\documentclass[aps,prb,twocolumn]{revtex4}
\usepackage{graphicx,subfigure}
\usepackage{bm}
\usepackage{amsfonts}
\usepackage{dsfont}
\usepackage{color}
\usepackage{enumerate}
\usepackage{bbold}

\begin{document}

\title{ Detection of Symmetry Protected Topological Phases in 1D}
\author{Frank Pollmann}
\affiliation{Max-Planck-Institut f\"ur Physik komplexer
Systeme, 01187 Dresden, Germany}
\author{Ari M. Turner}
\affiliation{University of Amsterdam, 1090 GL Amsterdam, The Netherlands}

\date{\today}

\begin{abstract}
A topological phase is a phase of matter which cannot be characterized by a local order parameter. It has been shown that gapped phases in 1D systems can be completely characterized using tools related to projective representations of the symmetry groups. We show how to determine the matrices of these representations in a simple way in order to distinguish between different phases directly.  From these matrices we also point out how to derive several different types of non-local order parameters for time reversal, inversion symmetry and $Z_2 \times Z_2$ symmetry, as well as some more general cases (some of which have been obtained before by other methods). Using these concepts, the ordinary string order for the Haldane phase can be related to a selection rule that changes at the critical point. We furthermore point out an example of a more complicated internal symmetry for which the ordinary string order cannot be applied.
\end{abstract}

\maketitle

\section{Introduction}
Phases of matter are usually identified by measuring a local order parameter. These order parameters reveal spontaneous symmetry breaking.\cite{landau37} In the $Z_2$ symmetric Ising model  we find for example an ordered and a disordered phase which can be distinguished by an order parameter which measures the magnetization. Over the last decades, it has been discovered that distinct quantum phases separated by quantum phase transitions can occur even when there is no local order parameter or spontaneous breaking of a global symmetry. These phases are usually referred to as ``non-trivial topological phases''.\cite{Wen89}   One of the simplest examples of a topological phase is the Haldane phase in quantum spin chains with odd integer spin.\cite{Haldane-1983,Haldane-1983a} 
By tuning various parameters, such as anisotropy terms, this state can be driven through a critical point. Yet on the other side of the critical point, there is no spontaneous symmetry breaking either.  A mystery then is to find some non-local order parameter or another property that changes
at the critical point. Such a property was first found by
considering an exactly solvable model
introduced by Affleck, Kennedy, Lieb, and Tasaki (AKLT)\cite{PhysRevLett.59.799,AKLT-CMP}. The ground state, the AKLT state, was later found to exhibit several unexpected properties, such as a non-local ``string order'' and edge states, which extend also to states within the same phase.\cite{DenNijsRommelse} 

It turns out that the topological phases in the Spin-1 chain can be understood in terms of ``fractionalization'' of symmetry operations at the edges\cite{Pollmann-2010,Turner-2010}, which is reflected in the bulk as well, by non-trivial degeneracies in the entanglement. In other words, different phases correspond to inequivalent projective representation of the symmetries present.
 These topological phases can be protected by any of the following symmetries: spatial inversion symmetry, time reversal symmetry or the $Z_2\times Z_2$ symmetry (rotations by $\pi$ about a pair of orthogonal axes).\cite{Chen-2011,Pollmann-2010} The same approach can be applied to phases with other symmetry groups--the phases can simply be classified by enumerating the possible types of projective representations of the appropriate group.  This  approach was then shown to give a complete procedure in one dimension, and elaborated in various directions.\cite{Chen-2011,Chen-2011a,Schuch-2011}

As the symmetry protected phases do (by definition) not break any symmetry, there exist no local properties in the bulk which can be measured to distinguish the phases. On the other hand, for certain cases, non-local order parameters have been derived to distinguish different symmetry protected phases. For example the string order mentioned above can be applied whenever the phases are stabilized by a $Z_2\times Z_2$ symmetry\cite{DenNijsRommelse} and some aspects of it have recently been generalized to other local symmetries.\cite{PerezGarcia-2008,Haegeman-2012} Furthermore, it has been found recently that string order can actually be observed experimentally: Endres et. al observed string order  in low-dimensional quantum gases in an optical lattice using high-resolution imaging.\cite{Endres-2011}

In this paper, we show how to convert the mathematical description of topological phases into a practical numerical procedure. This procedure  calculates the projective representation of the symmetries of a given state, starting from a matrix product state.  The projective representations can then be used to identify \emph{any} symmetry protected phase. However, this procedure is only practical when one has a matrix product representation of the state (or at least has a way of determining its entanglement spectrum). We therefore also discuss other non-local order parameters, generalizations of string order, which can be calculated from \emph{any} representation of the wave function using, e.g., Monte Carlo simulations or possibly even measured experimentally. In particular,  we revisit and generalize the
den Nijs and Rommelse string order for local symmetries\cite{DenNijsRommelse} and show that it works because of a selection rule that changes at the critical point. We conclude with an alternative order parameter for local symmetries (introduced by Ref. \onlinecite{Haegeman-2012}). The former type of order parameter is an easier way to identify phases for many symmetry groups.  However, we point out that there are some states with more complicated symmetry groups to which it does not apply, while Ref. \onlinecite{Haegeman-2012}'s order parameter always works. We furthermore explain non-local order parameters for the cases of inversion symmetry (which was introduced by Ref.~\onlinecite{Cen-2009}) and time-reversal symmetry.

This paper is organized as follows: We first  briefly review properties of MPS's and their transformation under symmetry operation in Sec.~\ref{mps}. In Sec.~\ref{nlo_mps}, we show how to distinguish MPS representations of different symmetry protected topological phases and present numerical results for a Spin-1 Heisenberg chain. In Sec.~\ref{nlo_wav}, we analyze non-local order parameters which can be calculated from any representation of the wave function. (The appendix gives an
example of a phase that cannot be identified using
the den Nijs-Rommelse string order, but can be identified
with the more general order of Ref. \onlinecite{Haegeman-2012}.) We finally summarize our results again in Sec~{\ref{con}}.

\section{Symmetries in Matrix product states}
\label{mps}

\subsection{Matrix product states}
\begin{figure}[tbp]
\includegraphics[width=8cm]{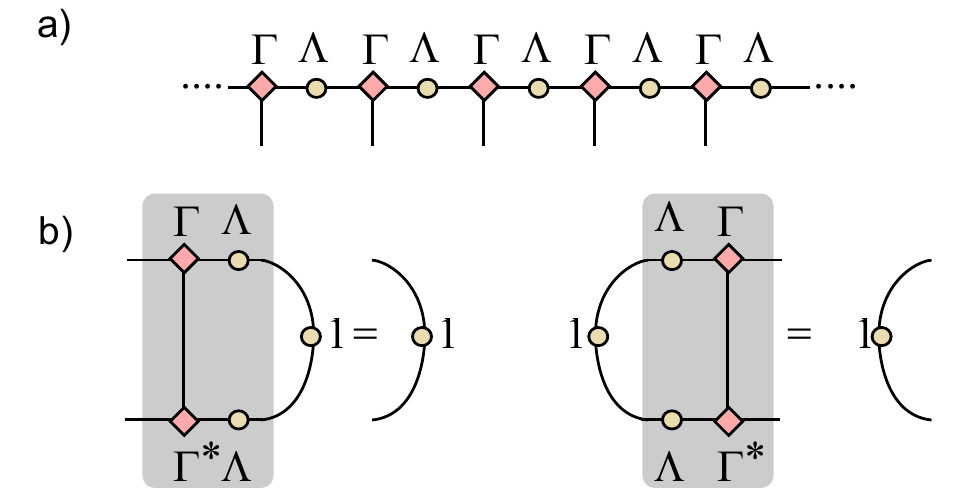}
\caption{(a) Diagrammatic representation of an  iMPS formed by the tensors $\Gamma$ and $\Lambda$. The horizontal line represents the bond indices $1\dots \chi$ and the vertical lines the physical indices $1\dots d$. (b) Condition for the MPS to be in the canonical form (i.e., the transfer matrices Eq.~(\ref{eq:transfer_r}) and Eq.~(\ref{eq:transfer_l}) have the identity as eigenvectors with eigenvalue one.}
\label{fig:mps}
\end{figure}

We  use a matrix product state (MPS) representation \cite{Fannes-1992} to understand and to define non-local order parameters for topological phases in 1D. We consider translationally invariant MPS's, using the framework contained in Ref.~\onlinecite{Vidal-2007}. A translationally invariant Hamiltonian on a chain of length $L$ has a ground
state that can  be written as the following MPS:
\begin{equation}
|\Psi \rangle = \sum_{j_1, \ldots, j_L} B^T A_{j_1} \ldots A_{j_L} B | j_1, \ldots ,j_L \rangle,  \label{eq:mps}
\end{equation}
where $A_{j}$ are $\chi \times \chi$ matrices, and $|j_k\rangle$
represents  local states at site $k$. The  $\chi \times 1$ matrix $B$ determines the boundary conditions. For most of this paper we consider the case of infinite chains and the boundary matrices can be ignored. Ground states of one dimensional, gapped
systems can be efficiently approximated by an MPS representation\cite{Hastings-2007, Gottesman-2009, Schuch-2008}, in the sense that the
value of $\chi $ needed to approximate the ground state
wave function to a given accuracy converges to a finite value as $
L\rightarrow \infty$. 

In this paper we follow Ref.~\onlinecite{Vidal-2007} and use infinite matrix-product states (iMPS's) for translationally invariant, infinite chains. In the iMPS representation, we write the matrices $A_{j}$ as a product of $\chi \times \chi $ complex matrices  $\Gamma _{j}$ and positive, real, diagonal matrices $\Lambda $ (see Fig.~\ref{fig:mps}a for a diagrammatic representation). The matrices $\Gamma_j $, $\Lambda$ can be chosen to be in a canonical form; that is, the transfer matrix
\begin{equation}
T_{\alpha \alpha ^{\prime };\beta \beta ^{\prime }}=\sum_{j}\Gamma _{j,\alpha\beta}\left(\Gamma _{j,\alpha^{\prime }\beta^{\prime }}\right)^{\ast}\Lambda _{\beta }\Lambda _{\beta ^{\prime }}
\label{eq:transfer_r}
\end{equation}
should have a right eigenvector $\delta _{\beta \beta ^{\prime }}(=\mathds{1})$ with eigenvalue $
\eta =1$ ($^\ast$~denotes complex conjugation). Similarly,
\begin{equation}
\tilde{T}_{\alpha \alpha ^{\prime };\beta \beta ^{\prime }}=\sum_{j}\Lambda _{\alpha }\Lambda _{\alpha ^{\prime }}(\Gamma_{j,\alpha ^{\prime }\beta ^{\prime }})^{\ast }\Gamma _{j,\alpha\beta }\label{eq:transfer_l}
\end{equation}
 has a left eigenvector $
\delta _{\alpha \alpha ^{\prime }}$ with $\eta =1$ (see Fig.~\ref{fig:mps}b for a diagrammatic representation). In this case, the diagonal matrix $\Lambda$ contains the Schmidt  values  $\lambda_{\alpha}$ for a decomposition into two half infinite chains,
\begin{equation}
|\Psi \rangle =\sum_{\alpha }\lambda _{\alpha }|\alpha L\rangle |\alpha
R\rangle ,  \label{schmidt}
\end{equation}
where $|\alpha L\rangle $ and $|\alpha R\rangle $ ($\alpha =1,\dots ,\chi $)
are orthonormal basis vectors of the left and right partition, respectively.
The states $|\alpha L\rangle$ and
$|\alpha R\rangle$ can be obtained by multiplying together all the
matrices to the left and right of the bond, e.g., if the broken
bond is between sites $0$ and $1$, the right Schmidt states are given by
\begin{equation}
 |\alpha R\rangle=\sum_{\{j_{k}\},k>0}\left[\prod_{l>0} 
\Gamma
_{j_{l}} \Lambda\right]_{\alpha\gamma} |j_1,j_2\dots\rangle.\label{eq:schm}
\end{equation}
Here, $
\gamma $ is the index of the row of the matrix; when the chain is infinitely
long, the value of $\gamma$ affects only an overall factor in the
wavefunction. Reviews of MPS's as well as the canonical form can be found in
Refs.~\onlinecite{PerezGarcia-2007,Orus-2008,Vidal-2007}.

Furthermore, we must require that our state is not a cat
state.  The condition turns out
to be that $\mathds{1}$ is the
\emph{only} eigenvector with eigenvalue $\left\vert \eta
\right\vert = 1$.\cite{PerezGarcia-2008}
The
second largest (in terms of absolute value) eigenvalue
$\epsilon_2$ determines the largest correlation length
\begin{equation}
 \xi = - \frac{1}{\log{|\epsilon_2|}} \label{eq:corr}.
\end{equation}

\subsection{Symmetry protected topological phases}

\begin{figure}[tbp]
\includegraphics[width=8cm]{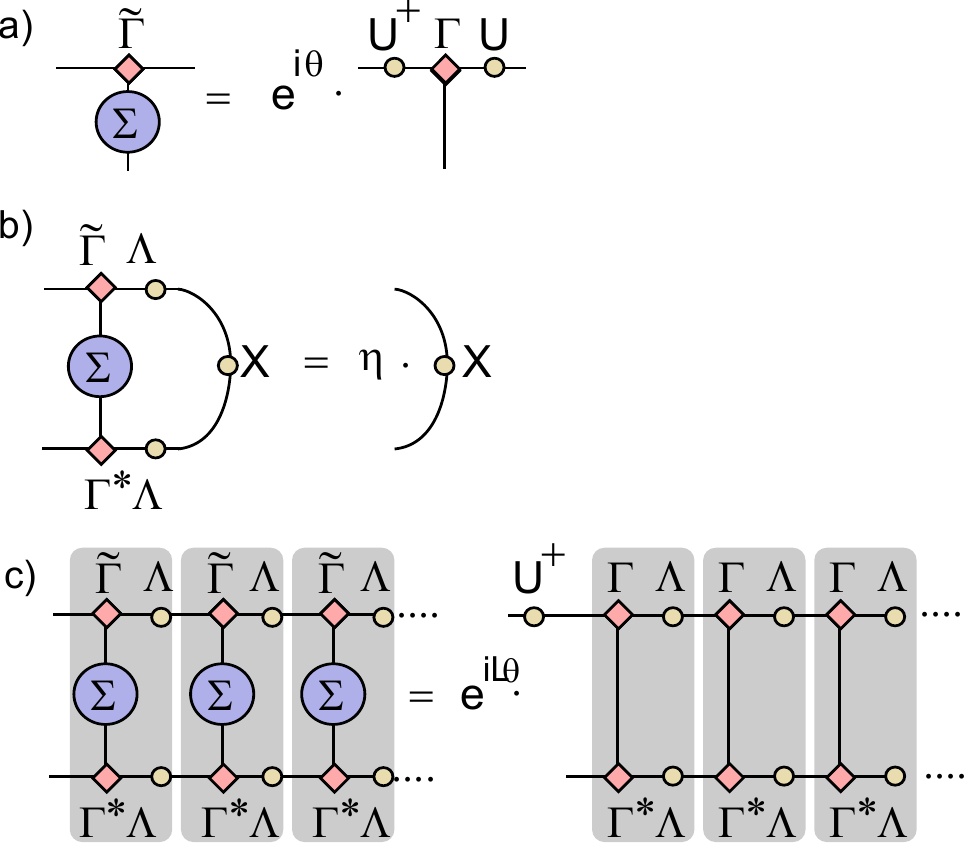} 
\caption{(a) Transformation of an iMPS which is invariant under an internal symmetry operation $\Sigma$. Here $\tilde{\Gamma}$ can be $\Gamma$, $\Gamma^*$ or $\Gamma^T$. (b) Eigenvalue equation $T^{\Sigma}X=\eta X$  for the generalized transfer matrix, where $\eta$ is the dominant eigenvalue. The upper part corresponds to the transformed wave function and the lower part to the original one.  We find  $|\eta|=$1 iff the state is symmetric under this transformation. (c) Overlap of Schmidt states $|\alpha R\rangle$ with its symmetry transformed partners. If the chain is assumed to be very long, the overlap can be expressed in terms of the eigenvector $X$ corresponding the largest magnitude eigenvector $|\eta|=1$ of the generalized transfer matrix (grey blobs). The right boundary yields an overall phase factor which we ignore here  (see text for details).}
\label{fig:trans}
\end{figure}

If a state $|\Psi \rangle $ is invariant under an \emph{internal symmetry}, which is represented in the spin basis as a unitary matrix $\Sigma
_{jj^{\prime }}$, then the $\Gamma_j $ matrices must transform under  $\Sigma
_{jj^{\prime }}$  in such a way that the product in Eq.~(\ref{eq:mps}) does not change (up to a phase). 
Thus the transformed matrices can be shown to satisfy\cite
{Pollmann-2010,PerezGarcia-2008}
\begin{equation}
\sum_{j^{\prime }}\Sigma _{jj^{\prime }}\Gamma _{j^{\prime }}=e^{i\theta}U^{\dagger }\Gamma _{j}U^{\vphantom{\dagger }}\text{,}  \label{trans}
\end{equation}
where $U$ is a unitary matrix which commutes with the $\Lambda $
matrices, and $e^{i\theta}$ is a phase factor(see Fig.~\ref{fig:trans}a for a diagrammatic representation). As
the symmetry element $g$ is varied over the whole group, a set of
phases and matrices $e^{i\theta_g}$ and $U_g$ results. The phases form a 1D representation (i.e., a character) of the symmetry group.  The matrices $U_g$ form a $\chi-$dimensional (projective) representation
of the symmetry group. A projective representation is like an
ordinary regular representation up to phase factors; i.e., if  $\Sigma^g\Sigma^h=\Sigma^{gh}$, 
then 
\begin{equation}
U_{g}U_{h}=e^{i\rho(g,h)}U_{gh}.
\label{eq:rho}
\end{equation}
The phases $\rho(g,h)$ can be used to classify different 
topological phases.\cite{Pollmann-2010,Schuch-2011,Chen-2011} Consider for example a
model which is invariant under a $Z_2\times Z_2$ symmetry of rotations $\mathcal{R}_x=\exp(i\pi S^x)$ and $\mathcal{R}_z=\exp(i\pi S^z)$. The phases for each spin rotation individually (e.g., $U_{x}^2=e^{i\alpha}\mathds{1}$) can 
be removed by redefining the phase of the corresponding $U$-matrix.  However, the representations of 
$\mathcal{R}_x\mathcal{R}_z$ and $\mathcal{R}_z\mathcal{R}_x$ can also differ by a phase, which it turns out must be $\pm 1$:
\begin{equation}
U_{x}U_{z}=\pm U_{z}U_{x}.
\label{eq:schur}
\end{equation}
I.e., the matrices either commute or anti-commute. This resulting  phase cannot be gauged away because the phases 
of $U_{x}$ and $U_{z}$ enter both sides of the equation in the same way. Thus we have two different classes of projective representations.

We can derive a similar relation to Eq. (\ref{trans}) for time reversal and
inversion symmetry.\cite{Pollmann-2010} For a time reversal transformation $\Gamma
_{j^{\prime }}$ is replaced by $\Gamma _{j^{\prime }}^{\ast }$
(complex conjugate) on the left hand side. In the case of
inversion symmetry $\Gamma _{j^{\prime }}$ is replaced by $\Gamma
_{j^{\prime }}^{T}$ (transpose) on the left hand side of Eq.
(\ref{trans}). In both cases we can distinguish the two different 
phases depending on whether  $U^{\vphantom{*}}_{\mathcal{I}}U_{\mathcal{I}}^*=\pm\mathds{1}$ and $U^{\vphantom{*}}_{\text{TR}}U_{\text{TR}}^*=\pm\mathds{1}$. Details on the classification of topological phases can be found in, e.g., Ref.~\onlinecite{Pollmann-2010,Chen-2011,Chen-2011a,Schuch-2011,Pollmann-2012}.

\section{Detecting  symmetry protected topological phases in MPS representations}
\label{nlo_mps}

The definitions in the previous section tell us exactly what kind of topological phases exist in 1D and how to classify them. It does, however, not give us a direct method to detect and distinguish different phases. In Ref.~\onlinecite{Pollmann-2010} it is pointed out that  topologically non-trivial phases must have degeneracies in the entanglement spectrum. However, this does
not distinguish among various non-trivial topological states (when there is more than 1). Furthermore, DMRG calculations sometimes produce states which have a degenerate entanglement spectrum for another reason (such as ``cat states" for a phase with broken symmetry). 

In this section we show to directly obtain the projective representations $U$, given that the ground-state wave function is given in the form of an iMPS, i.e., we have access to the $\Gamma_j,\Lambda$ matrices. These matrices can be conveniently obtained using various numerical methods, e.g.,  the infinite time-evolving block decimation (iTEBD) method.\cite{Vidal-2007} The iTEBD method is a descendent of the density matrix renormalization group (DMRG) method.\cite{White-1992} Once the algorithm has converged to the ground state, the   matrices are already in the desired canonical form. 

We will now explain how the $U$-matrices may be obtained by diagonalizing transfer matrices.\cite{PerezGarcia-2008}
First of all, we need to test if the iMPS is  invariant under certain symmetry operations, i.e, we require that $|\langle\psi|\tilde{\psi}\rangle|=1$ with $|\tilde{\psi}\rangle$ being the transformed state. This implies that a ``generalized'' transfer matrix 
\begin{equation}
T^{\Sigma}_{\alpha \alpha ^{\prime };\beta \beta ^{\prime }}=\sum_{j}\left(\sum_{j^{\prime }}\Sigma _{jj^{\prime }}\tilde{\Gamma} _{j^{\prime },\alpha\beta}\right)\left(\Gamma _{j,\alpha ^{\prime
}\beta ^{\prime }}\right)^{\ast}\Lambda _{\beta }\Lambda _{\beta ^{\prime }}
\label{eq:gtransfer}
\end{equation}
must have a largest eigenvalue $|\eta|=1$,
\begin{equation}
T^{\Sigma}_{\alpha\alpha';\beta\beta'}X_{\beta\beta'}=\eta X_{\alpha\alpha'};
\label{eq:X}
\end{equation} 
see also the diagrammatic representation in  Fig.~\ref{fig:mps}b. 
Here $\Sigma$ is an internal  symmetry operation and $\tilde{\Gamma}_j$ is equal to $\Gamma_j$,  $\Gamma^*_j$ or  $\Gamma^T_j$, depending on the symmetry of the system (the complex conjugate and transpose are required for time reversal and inversion, respectively).  If  $|\eta|<1$,  the overlap between the original and the transformed wave function decays exponentially with the length of the chain and $|\psi\rangle$ is thus not invariant. Given that $|\eta|=1$, the information about the symmetry protected topological phase of the  system is encoded in  the corresponding eigenvector $X_{\beta^{\prime}\beta}$. We will now see that $U$ is related
to $X$, specifically 
\begin{equation}
U_{\beta\beta'}=X^*_{\beta'\beta}.\label{eq:UX}
\end{equation}
 (For DMRG calculations of $U$, it is helpful to note that if the iMPS is not
obtained in the canonical form, we need to multiply by the inverse of the eigenstate of the transfer matrix Eq.(\ref{eq:transfer_r}).)

This convenient expression for finding $U$ can be understood from the symmetry transformation of the Schmidt states $|\alpha R\rangle$  defined in Eq.~({\ref{eq:schm}).  Figure \ref{fig:mps}c shows the overlap of the Schmidt states $|\alpha R\rangle$  with their transformed partners  $\Sigma|\tilde{\alpha} R\rangle$. The overlap corresponds to applying the generalized transfer matrix $T^{\Sigma}$ many times; hence only the dominant eigenvector $X_{\beta\beta^{\prime}}$ remains in the thermodynamic limit. On the other hand, we can apply the transformation  Eq.~(\ref{trans}) to each transformed matrix and see that  only the $U^{\dag}$ at the left end remains (Fig.~\ref{fig:mps}d). Using the fact that the matrices $\Gamma_j$, $\Lambda$ are chosen to be in the canonical form, we can read off that $(U^{\dag})_{\beta\beta^{\prime}}=X_{\beta\beta^{\prime}}$ (where we normalize $X$ such that $XX^\dagger=\mathds{1}$ and ignore a constant phase factor which results from the right end). Thus the $U$ matrices can be
obtained by finding the dominant eigenvector of the generalized transfer matrix, i.e., $T^{\Sigma}U^{\dagger}=e^{i\theta}U^{\dagger}$.
Once we have obtained the $U^{\dag}$ of each symmetry operation, we can read off in which phase the state is. 
Furtermore, we can directly see the block structure of the matrices which is discussed in Ref.~\onlinecite{Pollmann-2010}. 
\subsection{Example:  spin-1 chain}

\begin{figure}[tbp]
\includegraphics[width=6cm]{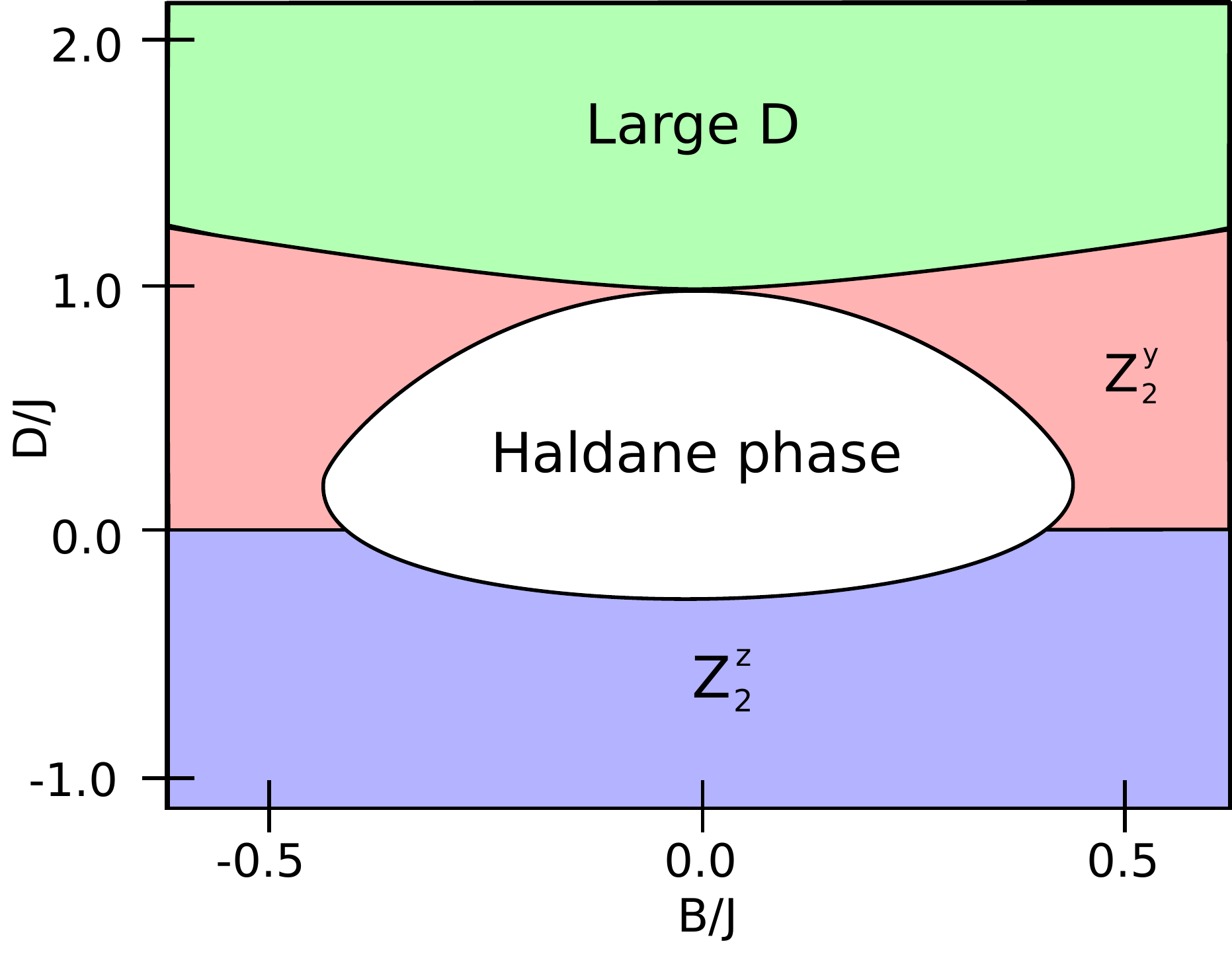}
\caption{Phase diagram of a spin-1 Heisenberg Hamiltonian in the presence of a single-ion anisotropy $D$ and a transverse magnetic field in $x$-direction with magnitude $B$.}
\label{fig:PD}
\end{figure}
In this section we show for an example how we can use this approach to distinguish different symmetric phases. We consider the  spin-1 model Hamiltonian
\begin{equation}
H=J \sum_{i} \vec{S}(i) \cdot \vec{S}(i+1)  + B \sum_i S^x(i) + D \sum_i [S^{z}(i)]^2. \label{H0}
\end{equation}
as a specific example in which symmetry protected topological phases occur. This model is invariant under translation, under spatial inversion
as well as under a combined $\pi$ rotation around the $y$-axis and
complex conjugation [$(S^x,
S^y,S^z)\rightarrow(S^x, -S^y, S^z)$].
The phase diagram  has been
studied in Ref.~\onlinecite{Gu-2009} and is shown in Fig.~\ref{fig:PD}.
The point $D=B=0$ is the Heisenberg point, around  which one finds the gapped Haldane phase. When $D$ increases, there is a transition into another phase which also does not break any symmetry. Even for $B\neq0$, there is a transition between these two phases (with an intervening phase). At large $D$, the 
phase is trivial and can be visualized by a state where all the sites are in the $\left\vert S^{z}=0\right\rangle $ state, hence
the phase containing the Heisenberg point seems to be a non-trivial topological phase.
 Furthermore, two antiferromagnetic phases $ Z_{2}^{y}$ and $Z_{2}^{z}$ with spontaneous
 non-zero expectation values of $\langle S^{y}\rangle $ and $\langle S^{z}\rangle $ are present,
 respectively. 

We now show how to use the method introduced in the previous section to  distinguish different symmetry protected topological phases in the spin-1 model. In the presence of a $Z_2\times Z_2$ symmetry ($B_x=0$), we can use the symmetry operations $\mathcal{R}^x=\exp(i\pi S^x)$ and $\mathcal{R}^z=\exp(i\pi S^z)$ (or alternatively any other pair of orthogonal spin rotations) to calculate the $\chi\times\chi$ matrices $U_{\mathcal{R}^x}$ and $U_{\mathcal{R}^z}$ as above.  From them
we can then define a quantity which distinguishes the different topological phases:
\begin{equation}
\mathcal{O}_{Z_2\times Z_2}=
\left\{
  \begin{array}{ c l }
     0&\mbox{if}\ |\eta_{R^x}|<1 \mbox{ or }  |\eta_{R^z}|<1\\
     \frac1{\chi}\mbox{tr}\left({U_{x}U_{z}U_{x}^{\dag}U_{z}^{\dag}}\right) &\mbox{if}\ |\eta_{R^x}|=|\eta_{R^z}|=1
  \end{array}\right..\label{eq:z2}
\end{equation}
Here $\eta_{R^x}$, $\eta_{R^z}$ are the largest eigenvalue of the generalized transfer matrix Eq.~(\ref{eq:transfer_r}). Thus $\mathcal{O}_{Z_2\times Z_2}=0$ if the state is not $Z_2\times Z_2$ symmetric and the two symmetric phases are distinguished by the properties of the $U$ matrices. If  $U_{x}$ and $U_{z}$ commute ($\mathcal{O}_{Z_2\times Z_2}=1$) the system is in a trivial phase (i.e, same class as a site factorizable state) and if they anti commute ($\mathcal{O}_{Z_2\times Z_2}=-1$), the system is in a nontrivial phase (i.e., the Haldane phase). We proceed in a similar way for the other symmetries. In the presence of inversion symmetry (i.e., $\Gamma \rightarrow \Gamma^T$), we define
\begin{equation}
\mathcal{O}_{\mathcal{I}}=
\left\{
  \begin{array}{ c l }
     0&\mbox{if}\ |\eta_{\mathcal{I}}|<1 \\
     \frac1{\chi}\mbox{tr}\left(U^{{\vphantom{*}}}_{\mathcal{I}}U_{\mathcal{I}}^*\right) &\mbox{if}\ |\eta_{\mathcal{I}}|=1.
  \end{array}\right.\label{eq:I}
\end{equation}
For time reversal symmetry the matrices transformation as $\Gamma_j\rightarrow \sum_{j^{\prime}}[\exp(i\pi S^y)]_{jj^{\prime}}\Gamma_{j^{\prime}}^*$ and the corresponding order parameter reads 
\begin{equation}
\mathcal{O}_{\text{TR}}=
\left\{
  \begin{array}{ c l }
     0&\mbox{if}\ |\eta_{\text{TR}}|<1 \\
     \frac1{\chi}\mbox{tr}\left(U_{\mathrm{TR}}^{\vphantom{*}}U_{\mathrm{TR}}^*\right) &\mbox{if}\ |\eta_{\mathrm{TR}}|=1
  \end{array}\right.\label{eq:TR}.
\end{equation}
These quantities behave  similarly to $\mathcal{O}_{Z_2\times Z_2}$, i.e.,  $\mathcal{O}_{\mathcal{I}/\text{TR}}=0$ if the symmetry is broken and $\mathcal{O}_{\mathcal{I}/\text{TR}}=\pm1$ for the two symmetric phases. 

\begin{figure}[tbp]
\includegraphics[width=8cm]{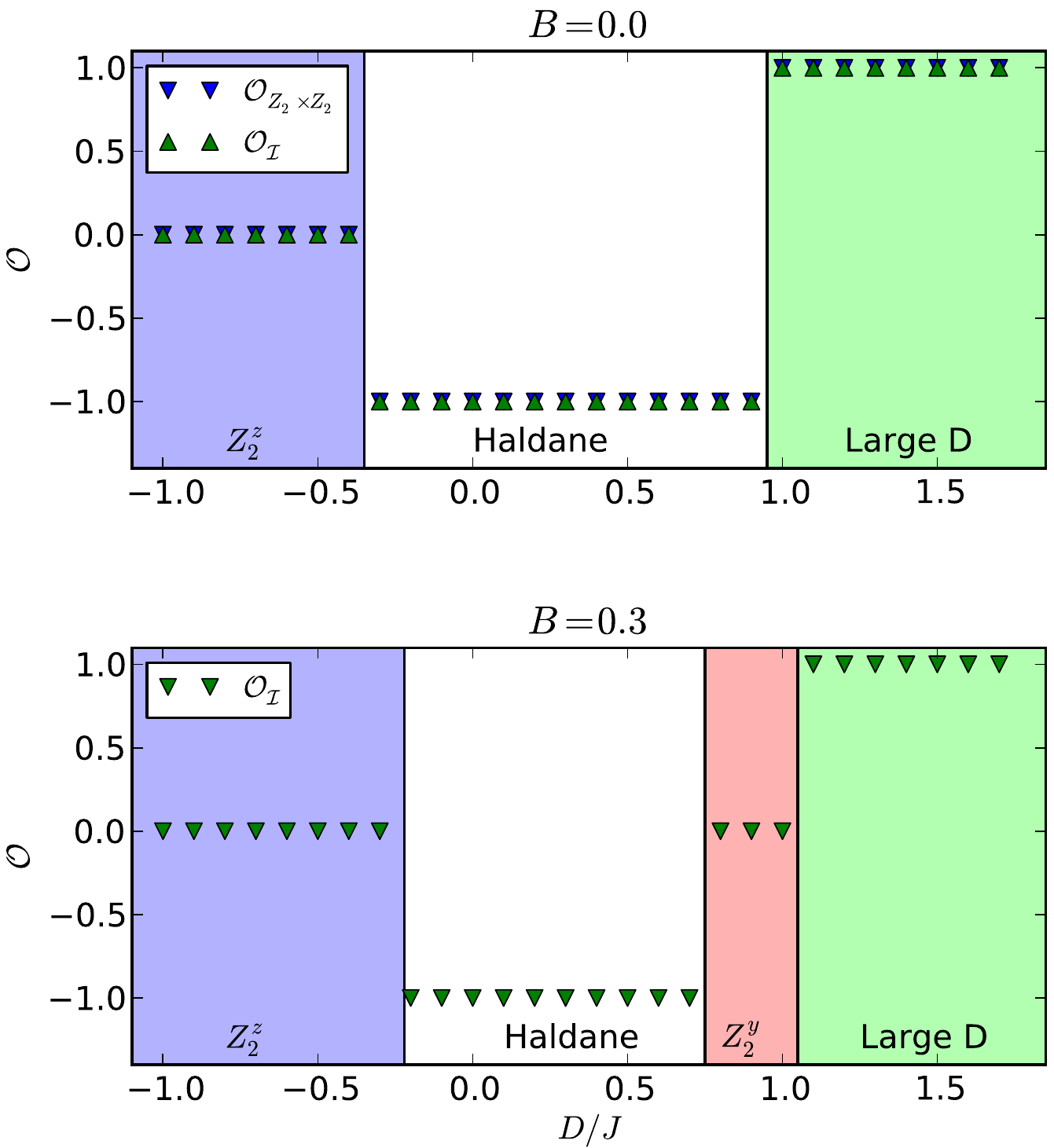}
\caption{Different phases of Hamiltonian (\ref{H0}) are distinguished by $\mathcal{O}_{Z_2\times Z_2}$ and $\mathcal{O}_{\mathcal{I}}$  (defined in the text). These quantities are equal to zero if the symmetry is broken and $\pm1$ is distinguishing the trivial and non-trivial phases.}
\label{fig:op}
\end{figure}

The procedure to calculate the quantities defined above is summarized by
Eqs. (\ref{eq:gtransfer}), (\ref{eq:X}), (\ref{eq:UX}) and the formulae Eqs. \ref{eq:z2}-(\ref{eq:TR}) for the appropriate symmetry. 
We use the iTEBD method\cite{Vidal-2007} to obtain the ground state of Hamiltonian Eq.~(\ref{H0}) in the desired canonical form. Then we construct the generalized transfer matrices Eq. (\ref{eq:gtransfer}) for the appropriate symmetry operations and find their largest eigenvalues $\eta$ with corresponding eigenvectors $X$ (Eq. (\ref{eq:X})) using sparse matrix diagonalization (the iTEBD algorithms breaks the translational symmetry, yielding two  matrices $\Gamma_j^{A/B}$ and thus we construct the transfer matrix using a two-site unit cell).  From $\eta$ and $X$ we determine the quantities defined in Eqs.~(\ref{eq:z2}) and (\ref{eq:I}). Figure 4 shows  the results  which  distinguish the phases clearly. Interestingly, the sharp distinctions between the phases can be achieved using MPS with rather small bond dimensions (we used MPS's with up to only $\chi=50$).

 \section{Non-local order parameters from a  wave function}
  \label{nlo_wav}

 In the previous section we showed how to detect different phases from an iMPS representation of the ground state. Now we derive expressions which can be evaluated when the wave function
is given in another form, for example, using Monte Carlo or possibly experimentally (as proposed by Ref.~\onlinecite{Endres-2011}). The basic idea is to find some operators on the physical Hilbert space which give
us some access to the $U$-matrices which live in the ``entanglement Hilbert space."
 \subsection{String order in the presence of internal symmetries}
 \label{sec:so}
 \begin{figure}[tbp]
\includegraphics[width=8cm]{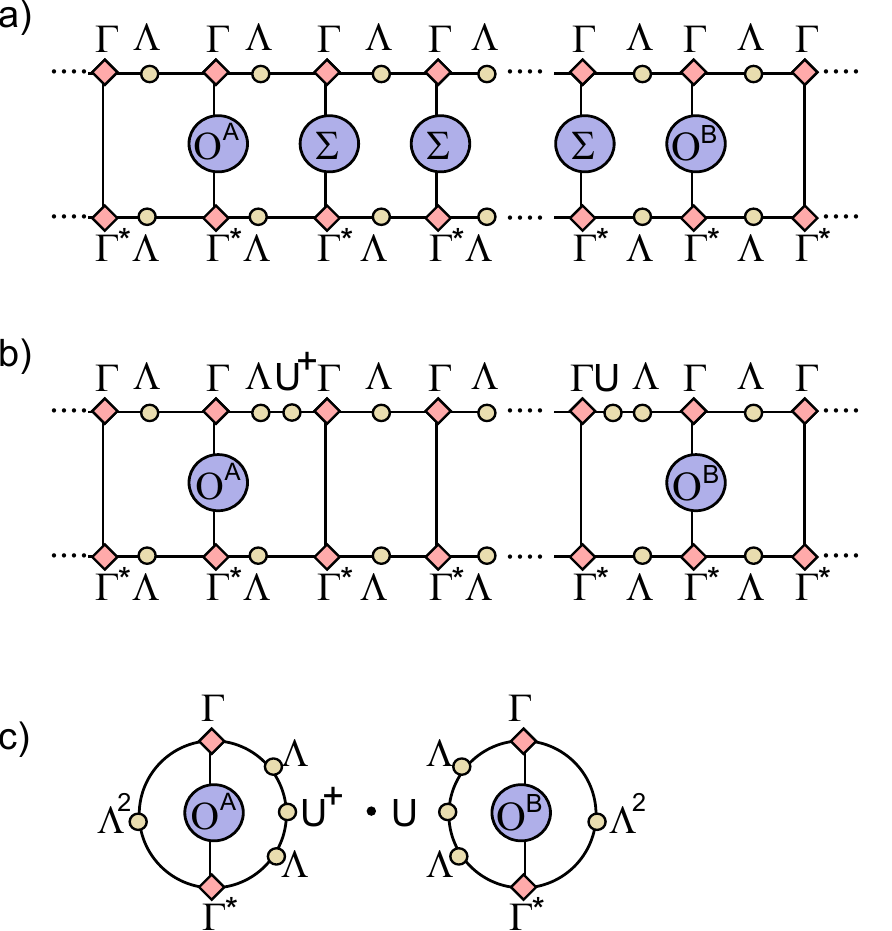} 
\caption{Diagrammatic derivation of the string order $\mathcal{S}$ for a wave function which is symmetric under an internal transformation $\Sigma$ and represented by an MPS in canonical form: (a) String order involving a segment of transformed sites terminated by operators $O^A$ and $O^B$. (b) The matrices $\Gamma_j$ transform according to Eq.~(\ref{trans}) and all matrices $U$ and $U^{\dagger}$ vanish except the ones at the edges
 (c) Using the properties of the transfer matrices (defined in the text), the expectation value in can be simplified for long segments.}
\label{fig:internal}
\end{figure}

Perez-Garcia et al.~\cite{PerezGarcia-2008} showed that the string order parameter, which was originally defined for $Z_2\times Z_2$ symmetric spin chains \cite{DenNijsRommelse} $\mathcal{S}_{\mathrm{str}}^{\alpha} \equiv \lim_{|j-k| \rightarrow \infty}\langle\psi_0| S^\alpha_j e^{i \pi \sum_{j \leq l < k} S^\alpha_l} S^\alpha_k|\psi_0 \rangle$, 
can be generalized for systems with other symmetry groups. The generalized form  for a state which is invariant under symmetry operations $\Sigma(k)$ reads:
\begin{equation}
\mathcal{S}(\Sigma,O^A,O^B)=\lim_{n\rightarrow\infty}\left\langle \psi_0\left|O^A(1)\left(\prod_{k=2}^{n-1}\Sigma(k)\right)O^B(n)\right|\psi_0\right\rangle.\label{eq:so}
\end{equation}
The non-vanishing of this expression for generic operators only
means that the state is symmetric, but does not distinguish
among topologically distinct states.

Nevertheless, we will now show that if the operators $O^A(1),O^B(n)$
are chosen appropriately, this order parameter can distinguish
some topological states.  However, it is not a complete characteristization
 since we give an example below showing that it does not  necessarily work in the presence of more complicated symmetries.

The most basic result about the string order correlator
Eq.~(\ref{eq:so}) is that {a phase must be symmetric under $\Sigma$
for the string order to be non-zero}. However, there is a second more
refined condition for when the string order is nonzero, which can
be used to distinguish between different symmetric phases. For example, the 
string order defined by $O^A=O^B=S^z$ vanishes in the large $D$ phase. Why does this occur
even though the state is symmetric?  The same string order is non-zero
in the Haldane phase, hence it seems to be connected to the topological
order of the phase, as we will show now.

Intuitively, the string order corresponds to calculating the overlap between the wave function with $\Sigma$ applied to $L$ consecutive sites and
the wave function itself.  Since $\Sigma$ is a symmetry of the wave function, it does not change anything in the bulk and the overlap should not vanish, generically speaking.  A diagrammatic representation of the string order is shown in Fig.~\ref{fig:internal}a. We represent the symmetry that is sandwiched in the middle using Eq. (\ref{trans}), i.e., $\sum_{j^{\prime }}\Sigma _{jj^{\prime }}\Gamma _{j^{\prime }}=e^{i\theta}U^{\dagger }\Gamma _{j}U^{\vphantom{\dagger }}$. Ignoring the overall phase factor $e^{in\theta}$, we obtain the expression shown in Fig.~5b. 
If $n$ is large, the part in between the $U^\dagger$ and $U$ is a product of  orthogonal Schmidt states of the segment yielding a scalar product of  delta functions $\delta_{\alpha\alpha'}$ on the left and $\delta_{\beta\beta'}$ on the right, yielding Fig.~\ref{fig:internal}c. 
That is, the string order is equal to the product $(\mathrm{tr}\ \Lambda\bar{O}^A\Lambda U^\dagger)(\mathrm{tr}\ \Lambda\bar{O}^B\Lambda U^T)$ (where
$\bar{O}^A_{\alpha'\alpha}=\langle \alpha'L|O^A|\alpha L\rangle$, or explicitly $\sum_\beta\frac{\lambda_\beta^2}{\lambda_\alpha\lambda_{\alpha'}}T^{O^A}_{\beta\beta,\alpha\alpha'}$ with the generalized transfer matrix $T^{O^A}$
as defined in Eq. (\ref{eq:gtransfer}), and where $\bar{O}^B_{\alpha'\alpha}=\langle\alpha' R|O^B|\alpha R\rangle=\sum_\beta T^{O^B}_{\alpha\alpha',\beta\beta}$).
This expression is nonzero unless one of the two factors is equal to zero.
Thus, the string order is generically non-zero in a symmetric phase.

Whether the factors vanish depends on the symmetry of the operators $O^A(1),O^B(n)$ and can be seen as a selection rule for string order. Such selection rules
exist only in the presence of additional symmetry. Thus, suppose that there are two symmetry operations $\Sigma^a$ and  $\Sigma^b$ which commute but $U^{\vphantom{\dagger}}_{b}U^{\vphantom{\dagger}}_{a}U_{b}^{\dagger} = e^{i\phi}U_{a}$.  We consider the string correlator $\mathcal{S}(\Sigma^a,O^A,O^B)$, and focus on the left side of it.
The operator $O^A$ can be chosen as having a particular quantum number under $\Sigma^b$, i.e., $\Sigma^b O^A (\Sigma^b)^{\dagger}=e^{i\sigma}O^A$. Then a short calculation shows that $\bar{O}^A$ transforms
in the same way under $U_b$, $U_b\bar{O}^AU_b^{\dagger}=e^{i\sigma}\bar{O}^A$. It follows that 
\begin{eqnarray}
\mathrm{tr}\ \Lambda \bar{O}^A\Lambda U^{\dagger}_{a} &=& \mathrm{tr}\ (U_b\Lambda \bar{O}^A\Lambda U^{\dagger}_{a}U_{b}^\dagger) \nonumber\\ &=&e^{i(\sigma-\phi)}\mathrm{tr}\ \Lambda\bar{O}^A\Lambda U^{\dag}_{a}.\end{eqnarray}

Thus we obtain a string order selection rule: the string order parameter vanishes if $\sigma\neq \phi$.  Without the second
symmetry $\Sigma^b$, the string order would not vanish.  Hence a nonzero string order in a state (though intuitively surprising) is actually not so unusual; it is the \emph{vanishing} of a string order that is the signature of a topological phase.  To summarize, the second criterion for
the string order is that \emph{$\sigma=\phi$ or else the string order vanishes.}
 
The string order for the spin-1 Heisenberg chain can, for example,  be derived simply in this way. Consider the Heisenberg chain with the symmetries $\mathcal{R}^x=\exp(i\pi S^x)$ and $\mathcal{R}^z=\exp(i\pi S^z)$.
Then the selection rule implies that the string order vanishes
in the trivial phase if one of the operators $O^A$, $O^B$
is odd under $180^\circ$ flips about the $x$ axis.  The string
order vanishes in the nontrivial phase if one of these operators is 
\emph{even} (since $U^z$ is odd under flips about the $x$-axis in this phase).  Thus, $\langle \psi_0|\mathds{1}\left( \prod_{k=2}^{n-1} \mathcal{R}^z(k)\right) \mathds{1}|\psi_0\rangle$ vanishes
in the nontrivial ($\phi=\pi$) phase and $\langle \psi_0|S^z(1)\left( \prod_{k=2}^{n-1} \mathcal{R}^z(k)\right) S^z(n)|\psi_0\rangle$
does not, while the situation is reversed in the trivial ($\phi=0$)
phase.  This is different than ordinary ordering
transitions as, e.g., for the Ising
model, where even operators have long-range correlations
in both phases.

This approach may be used to give an order parameter that is sensitive
to certain phase factors, those of the form $U^{\vphantom{\dagger}}_aU^{\vphantom{\dagger}}_bU_a^{\dagger}U_b^\dagger=e^{i\phi}$
for commuting symmetries. In order to determine $\phi$ systematically,
find test operators $O$  with each possible transformation
under $\Sigma^b$, and then see which of these has a non-zero string
correlation. In more detail, note first that
$\phi=\frac{2\pi k}{r}$ where $r$ is the order of $\Sigma_b$ and where $k$
is some integer, and thus
finding $\phi$ is equivalent to finding $k$.  We can then
choose ``test operators"
that are powers of a single operator $O_1$ that transforms as
 $(\Sigma^b)^{\dagger} O_1\Sigma^b=e^{\frac{2\pi i}{r}}O_1$. For $0\leq l\leq r-1$ calculate the string order
$\mathcal{S}_l=\mathcal{S}(\Sigma^a,O^A,O^B)$ with $O^A=(O_1)^l$, translated to the left end
of the segment, and $O^B=(O_1^\dagger)^l$, translated to the right end. The result
will be nonzero only for one value of $l$, namely $l= k$. 

In general, the possible phases of a system with a given symmetry group can be classified by finding all the consistent 
phase factors for a projective representation (see Eq. (\ref{eq:rho})). A phase can thus be identified
by measuring the gauge-invariant combinations
of these phase factors.
The procedure just given works for phase factors that arise from a pair
of symmetries that commute in the original symmetry group. However, for complicated groups, these might not be the only
parameters that one needs.   If $\Sigma^a$ and $\Sigma^b$ do not commute, e.g. $\Sigma^a\Sigma^b(\Sigma^a)^{-1}(\Sigma^b)^{-1}=\Sigma^x$ (another symmetry),
then there can be a phase in the projective representation, $U^{\vphantom{-1}}_aU^{\vphantom{-1}}_bU_a^{-1}U_b^{-1}=e^{i\phi_1}U_x^{\vphantom{-1}}$.
This phase cannot be detected using the string-order selection rule, but it also does not matter since
it is not gauge invariant: it can be absorbed into $U_x$. To give an example of a \emph{gauge invariant}
phase that cannot be detected by a selection rule, we need to involve more symmetries.  Suppose there
is another pair of symmetries $\Sigma^c$ and $\Sigma^d$ with the same commutator, i.e., $\Sigma^c\Sigma^d(\Sigma^c)^{-1}(\Sigma^d)^{-1}=\Sigma^x$.
Then in the projective representation, $U^{\vphantom{-1}}_cU^{\vphantom{-1}}_dU_c^{-1}U_d^{-1}=e^{i\phi_2}U_x^{\vphantom{-1}}$. Either $\phi_1$
or $\phi_2$ may be absorbed into $U_x$, but not both. In fact, we can write:
\begin{equation}
U^{\vphantom{-1}}_aU^{\vphantom{-1}}_bU_a^{-1}U_b^{-1}U^{\vphantom{-1}}_dU^{\vphantom{-1}}_cU_d^{-1}U_c^{-1}=e^{i(\phi_1-\phi_2)}\mathds{1}\label{eq:alphabet}
\end{equation}
 Thus, $e^{i(\phi_1-\phi_2)}$ is an example
of a phase factor that is gauge invariant but cannot be detected by the string order just developed. (Appendix \ref{app1} fleshes out the details of this example.)
This phase factor \emph{can} be detected by the general approach we started with, of diagonalizing transfer matrices
to find the $U$'s, and then just calculating the appropriate products of them. 

At the end of the next section, we will describe another
type of non-local order parameter that is sensitive to these
more complicated local phase factors. Similar types of
order parameters can also be used to identify phases protected by time
reversal or inversion symmetry.
In fact, these types of order parameter can detect any gauge-invariant phase-factor, so they give a complete way to
determine what type of symmetry-protected topological order a system has.

 \subsection{\label{sec:phasestring}Non-local order parameters that measure the phase factors}
Phases that are protected by inversion symmetry, time-reversal symmetry or more complicated internal symmetries (see example above) cannot be detected using the selection rules. However, 
there is another type of non-local order parameter (for example introduced by Ref.~\onlinecite{Cen-2009} for inversion). The important thing about this parameter is not whether
it vanishes, but what its complex phase is.  The phase is simply
equal to a gauge-invariant phase in the projective representations, such as the sign of $U_{\mathcal{I}}U_{\mathcal{I}}^*$.

\paragraph*{Inversion symmetry.} In this case we can define an order parameter by simply reversing a part of the chain
with an even length and then calculating the overlap:
\begin{equation}
\mathcal{S}_{\mathcal{I}}(2n)=\langle \Psi|\mathcal{I}_{1,2n}|\Psi\rangle\label{strinv}
\end{equation} where $\mathcal{I}_{1,2n}$ is the inversion on the segment from $1$ to $2n$. This expectation value can be evaluated using, e.g.,  Monte Carlo methods and it distinguishes the two possible  symmetric phases by
\begin{equation}
\lim_{n\rightarrow\infty}\mathcal{S}_{\mathcal{I}}(2n)=\pm  \mathrm{tr}\Lambda^4,
\end{equation}
where  $\Lambda$ is a diagonal matrix  which contains the Schmidt values (as defined in Sec.~\ref{mps}). The \emph{sign} of this quantity
determines which of the two inversion-protected phases the chain is in. 
$\mathcal{S}_{\mathcal{I}}(2n)$ can be described by the
following thought experiment: Form pairs of sites
that are symmetric about the midpoint of the segment, and perform a measurement of the parity $P_k=\pm 1$ of the state of each pair.
Then $\mathcal{S}_{\mathcal{I}}(2n)$ is $\langle \prod_{k=1}^n P_k\rangle$. A non-zero value for $\mathcal{S}_I$ means
that there is a non-local correlation according
to which the number of odd pairs is more likely to be either an even or odd number, even when the chain is very long.
The effect is not so easy to see experimentally, since independent errors in measuring individual pairs will make even
and odd numbers equally likely.

\begin{figure}[tbp]
\includegraphics[width=8cm]{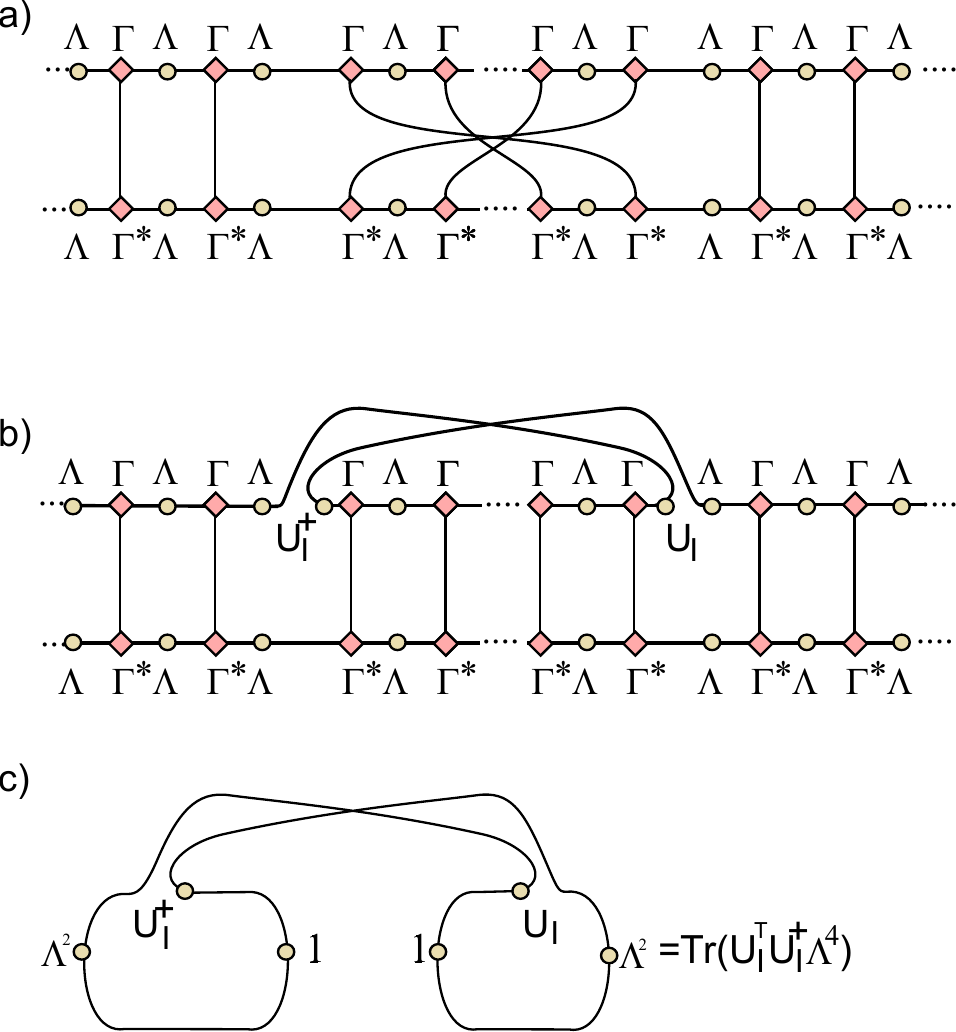} 
\caption{Diagrammatic derivation of the string order $\mathcal{S}_{\mathcal{I}}$ for a wave function which is inversion symmetric and represented by an iMPS in canonical form: (a) Overlap of a wave function with the wave function for an infinite chain in which a segment of $n$ sites has been inverted. (b) The overlap can be untwisted by reversing the segment using the unitaries $U_{\mathcal{I}}$. (c) For large $L$ and $n$, the expression can be  can be simplified by keeping only the largest magnitude eigenvector of the transfer matrix $T$, yielding $\mathcal{S}_{\mathcal{I}}$. }
\label{fig:cable}
\end{figure}

We now derive the string order $\mathcal{S}_{\mathcal{I}}(2n)$ formally using the iMPS representation together with the identities defined in Sec.~\ref{mps}. The result of reversing a segment
and taking the overlap, shown in Fig. \ref{fig:cable}a, is that the segments attaching the two chains to
each other get twisted. These may be untwisted by reversing the orientation of the segment on the top level
of the chain, at the expense of introducing a twist in \emph{it} (see Fig. \ref{fig:cable}b); 
the explicit calculation uses the relationship $\Gamma_{j}^T=e^{i\theta}U_{\mathcal{I}}^\dagger\Gamma^{\vphantom{\dagger}}_jU^{\vphantom{\dagger}}_{\mathcal{I}}$,
and so factors of $U_{\mathcal{I}}$ appear in the diagram. Now, each of the ladders is a product of several copies of the transfer matrix $T$,
and so, if $n$ is large, it becomes a projection onto the largest eigenvector of $T$, which is $\delta_{\alpha\beta}$ (see Eq.~(\ref{eq:transfer_r})), allowing
the diagram to be simplified again (see Fig. \ref{fig:cable}c and Fig. \ref{fig:cable}d). Reading along the loop in this figure gives the value of the string
order parameter:
\begin{equation}
\lim_{n\rightarrow\infty}\mathcal{S}_{\mathcal{I}}(22n)=\mathrm{tr}\ \Lambda U_{\mathcal{I}}^T\Lambda^2U_{\mathcal{I}}^\dagger\Lambda= \mathrm{tr} U_{\mathcal{I}}^TU_{\mathcal{I}}^\dagger\Lambda^4=\pm \mathrm{tr}\Lambda^4.
\end{equation}
\begin{figure}[tbp]
\includegraphics[width=8cm]{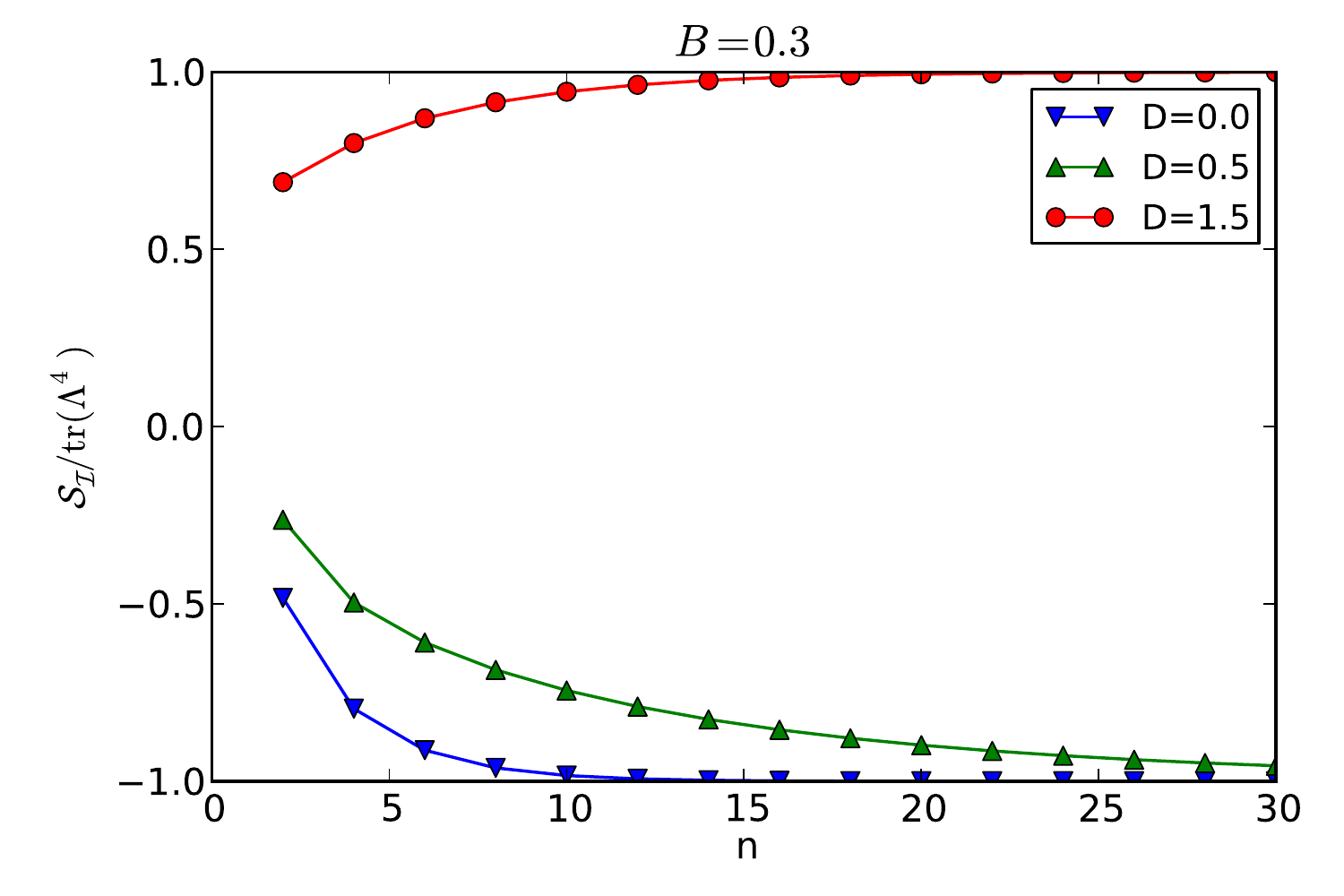} 
\caption{The non-local order parameter $\mathcal{S}_{\mathcal{I}}$ distinguishing  the Haldane ($D=0.0$, $D=0.5$) and large-D phase ($D=1.5$) in the presence of inversion symmetry.}
\label{fig:SI}
\end{figure}
Note that the  line goes backwards through $U_{\mathcal{I}}$ so it is transposed. Unlike the ordinary string order (which vanishes in one phase), it is the sign of this parameter that distinguishes among phases. (The factors of $e^{i\theta}$ cancel.)

As a specific example, we calculated $\mathcal{S}_{\mathcal{I}}$ for the spin-1 Heisenberg chain (\ref{H0}) in the presence of a finite transverse field. In this case, the Haldane phase is stabilized by  inversion symmetry. The results, which have been obtained using the iTEBD algorithm, are shown in Fig.~\ref{fig:SI}. The order parameter shows a clear distinction between the two phases. As we approach the phase transition, the correlation length gets longer and we have to make the segment longer to see the convergence of $\mathcal{S}_I/\mathrm{tr}\ (\Lambda^4)$ to $\pm 1$. 

\begin{figure}[bp]
\includegraphics[width=8cm]{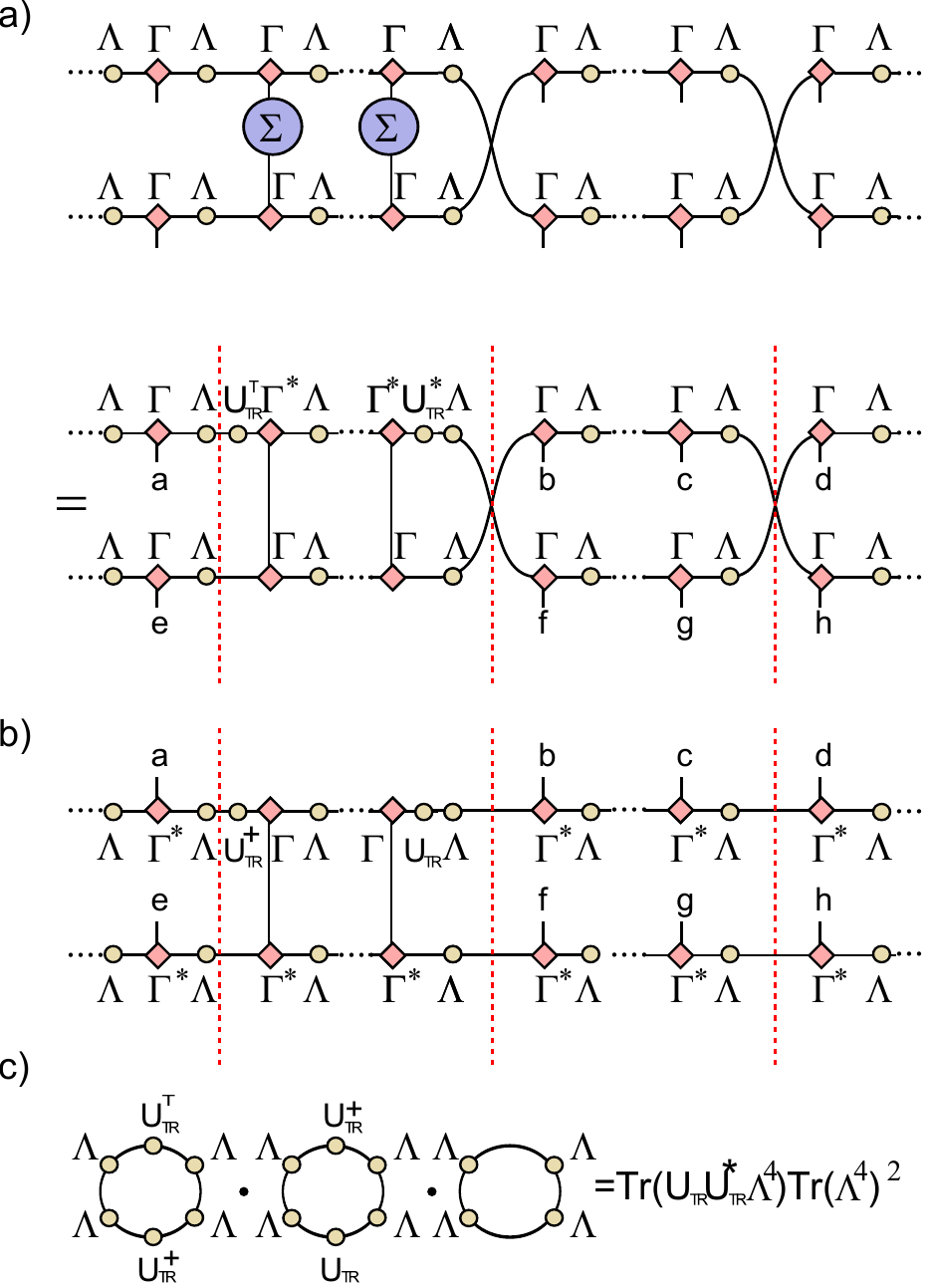} 
\caption{Diagrammatic derivation of the string order $\mathcal{S}_{\mathrm{TR}}$ for a wave function which is time-reversal symmetric and represented by an MPS in canonical form: (a) Representation of the wave function $d^{\frac{n}{2}}\langle R_{1n}|\mathrm{Swap}_{n+1,2n}|\psi_2\rangle$ containing three domain walls, where  
$\Sigma=\exp(i\pi S^y)/\sqrt{d}$  and $d$ is the physical dimension. The unitary $U_{\mathrm{TR}}$ represents $\Sigma$ and the complex conjugation of the segment in the upper row. (c) Representation of the wave function $d^{\frac{n}{2}}\langle\psi_2|R_{1n}\rangle$, containing two domain walls. (d) Contraction of the two wave functions  over open indices with the same letter [a-h] and keeping only the largest magnitude eigenvector of the transfermatrix $T$ yields for larger $m$,$n$ the order parameter $\mathcal{S}_{\mathrm{TR}}$. The dashed red lines indicate the positions of the domain walls.\label{fig:tr}}
\end{figure}

\paragraph*{Time Reversal.} A more complicated expectation value can be used to distinguish between
phases protected by time reversal symmetry. This order parameter is more subtle to devise
because there is no way to apply time-reversal to just a portion of a system (as there is for inversion symmetry) since time-reversal is anti-unitary.
For example, consider the state:
\begin{equation}
\frac1{\sqrt{2}}[(|1\rangle)|0\rangle+(i|0\rangle)|1\rangle]=\frac1{\sqrt{2}}[|1\rangle)|0\rangle+|0\rangle(i|1\rangle)]
\end{equation}
 formed
from two q-bits. One cannot apply time-reversal symmetry to just the first q-bit in a unique way, since
the result
of applying $\mathcal{T}$ to just the first atom comes out differently for the two ways of grouping the factors.
However, there is a way to express the overlap on the full chain $\langle \Psi|\mathcal{T}|\Psi\rangle$ without using antiunitary operators, and this is the starting point for a non-local order parameter. Consider for example a single spin $S=1$ with $j=-1,0,1$. Then
\begin{equation}
\kappa=\langle \Psi|\mathcal{T}|\Psi\rangle=\sum_{j,j^{\prime}} [e^{i\pi S^y}]_{j^{\prime}j}\Psi_j^*\Psi_{j^{\prime}}^*,
\end{equation}
is
not the ordinary expectation value of $e^{i\pi S^y}$ because both factors of $\Psi$ have complex conjugates.  To relate this to an expectation value, let us take two copies of the
system and introduce the two states, 
\begin{eqnarray}
|\Psi_2\rangle&=&|\Psi\rangle\otimes|\Psi\rangle\\
 |R\rangle&=&\frac{1}{\sqrt{3}}\sum_j [e^{i\pi S^y}]_{jj^{\prime}}|j\rangle\otimes |j'\rangle,
\end{eqnarray}
so that $\kappa=\sqrt{3}\langle \psi_2|R\rangle$. The phase of $\kappa$ is not well-defined, since
it depends on how one chooses the phase of $\psi$ so $|\kappa|^2=3\langle\psi_2\left(|R\rangle\langle R|\right)|\psi_2\rangle$ is a more
useful quantity. This can be related to an experiment where one takes two unentangled copies of the system and measures their state in a basis including
$R$.  The probability that the state is $R$ is then given by $|\kappa^2|/3$.

Now the generalization of $|\kappa|^2$ to an entire chain is useful for testing whether time-reversal
is broken spontaneously, but it does not help to distinguish between different phases.
For that, an operator has to be applied over part of the chain in order
to create ``domain walls" which depend on $U_\mathcal{T}$.  Therefore, we introduce an entangled state on just $n$ sites,
\begin{equation}
|R_{1n}\rangle=\prod_{k=1}^n \left(\frac{1}{\sqrt{3}}\sum_{j_k} [e^{i\pi S^y}]_{j_kj_k^{\prime}}|j_k\rangle\otimes |j_k^{\prime}\rangle\right).
\end{equation}
To define an order parameter that can distinguish different symmetric phases, we also have to introduce a swapping operator (defined similarly to Ref.~\onlinecite{Isakov-2011}).
Let $\mathrm{Swap}_{n+1,2n}$ swap the parts of the chains between $n+1$ and $2n$. Then we find that
\begin{eqnarray}
\mathcal{S}_{\mathrm{TR}}(n)&=&d^n\langle\psi_2|\left(|R_{1n}\rangle\langle R_{1n}|\right)\mathrm{Swap}_{n+1,2n}|\psi_2\rangle\nonumber\\
&=&\pm(\mathrm{tr}\Lambda^{4})^3.
\label{eq:time}
\end{eqnarray}
with $d$ being the local dimension of the Hilbert space.The swapping operator $\mathrm{Swap}_{n+1,2n}$   is introduced because, without it, $\langle \Psi_2\left(|R_{1n}\rangle\langle R_{1n}|\right)|\Psi_2\rangle$ does not depend on the sign of $U_\mathcal{T}U_\mathcal{T}^*$ (it is clearly positive).  Multiplying by $\mathrm{Swap}_{n+1,2n}$ 
makes it possible to isolate the phase $e^{i\phi_\mathcal{T}}$.

Figure~\ref{fig:tr} shows how to work out the order-parameter $\mathcal{S}_{\mathrm{TR}}$. The expectation value is evaluated in two parts:
First, we calculate $d^{\frac{n}{2}}\langle R_{1n}|\mathrm{Swap}_{n+1,2n}|\Psi_2\rangle$ (Fig.~\ref{fig:tr}a) and then $d^{\frac{n}{2}}\langle\Psi_2|R_{1n}\rangle$ (Fig.~\ref{fig:tr}b).  Since $|R_{1n}\rangle$ extends only over
$n$ sites, these are partial inner products, giving a wave-function in which $n$ spins have been removed. 
The short sticks coming out of the other sites represent the sites that have not been contracted yet.
Next, we transform Fig.~\ref{fig:tr}a using Eq.~(\ref{trans})
in the conjugate form $\sum_{j^{\prime}}[e^{i\pi S^y}]_{jj^{\prime}}\Gamma_{j^{\prime}}=U_{\mathrm{TR}}^T\Gamma^*_jU_{\mathrm{TR}}^*$ and  take the overlap between Figs. \ref{fig:tr}a and \ref{fig:tr}b, by contracting the short sticks with one another.  There will be three
``domain-wall" regions that we have to concentrate on (the bonds
between $0,1$; $n,n+1$ and $2n,2n+1$); everything else can be simplified by replacing the ladders by projections onto
the identity (in the same way as we have done several times before). 
The three ``domain walls" can be replaced by the product of loops in Fig. \ref{fig:tr}c.  
Contracting these expressions gives $\lim_{n\rightarrow\infty}\mathcal{S}_{TR}(n)=\left(\mathrm{tr}\ U^{\dagger}_{\mathrm{TR}}\Lambda^4 U_{\mathrm{TR}}\right)\left(\mathrm{tr}\ U_{\mathrm{TR}}^*\Lambda^4U^{\vphantom{\dagger}}_{\mathrm{TR}}\right)\left(\mathrm{tr}\ \Lambda^4\right)$, which is equal to Eq. (\ref{eq:time}).  (If the swap had not been included, the domain wall on
the right of the region contracted with $R$ would be proportional to $U_{\mathrm{TR}}^\dagger U^{\vphantom{\dagger}}_{\mathrm{TR}}=\mathds{1}$. The swap reverses the orientation of one of the
paths so that $U_{\mathrm{TR}}^*$ appears instead of $U_{\mathrm{TR}}^\dagger$, yielding the desired phase factor.)
This string order is nonzero in both the time-reversal-protected phases,
but when it is negative, the phase is non-trivial, just as for the 
inversion-symmetry. 
 
\paragraph*{Combinations of Multiple Local Symmetries}
A similar type of string order expectation value can also be used to identify more tricky phase factors, like in Eq. (\ref{eq:alphabet}). This type of string order was introduced by Ref.~\onlinecite{Haegeman-2012}.
To be general, note that there is
a gauge invariant phase
any time there is a sequence of symmetries $a_j$ which can be multiplied together
to give the identity in more than one way, $a_1a_2\dots a_m=a_{k_1}a_{k_2}\dots a_{k_m}=\mathds{1}$ (where the indices $k_1,\dots k_m$ are a permutation of $1\dots m$).
When these symmetries are replaced by the $U$'s, a phase
factor appears, $U_{k_1}\dots U_{k_m}=e^{i\phi}U_{1}\dots U_{m}$,
and the phase factor is gauge invariant.
For example, in Eq. (\ref{eq:alphabet}), the symmetries can be multiplied
together in the order $a^{\vphantom{-1}}b^{\vphantom{-1}}a^{-1}b^{-1}d^{\vphantom{-1}}c^{\vphantom{-1}}d^{-1}c^{-1}$ which gives the identity by
assumption or each symmetry can
be grouped with its inverse, which also cancels to $\mathds{1}$.
We will now see that such phase factors can be identified by taking multiple chains and applying symmetries and permutations
to them. The point we would most like to make about this
order parameter is that it succeeds at identifying phase factors that the string order selection rule fails to detect. 
In fact, it gives a complete way to distinguish between  topological states, because Schur showed that every gauge-invariant phase factor in a projective representation has this form, see Appendix \ref{app:schurmultiplier}.

The string order is illustrated in Fig. \ref{fig:musicalstairs} for $m=4$.
\begin{figure}
\includegraphics[width=.5\textwidth]{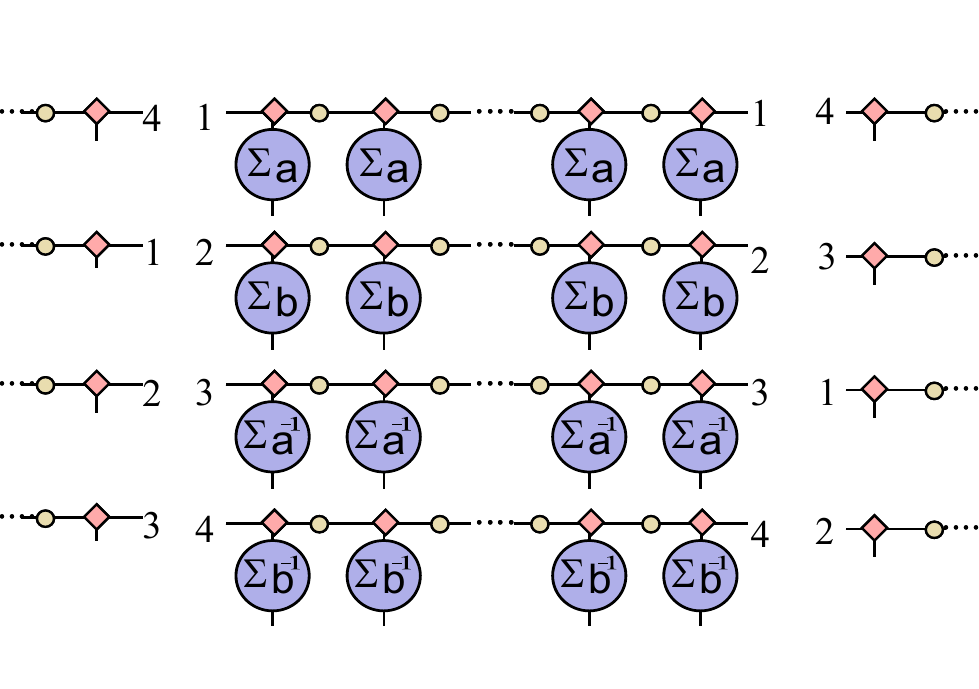}
\caption{\label{fig:musicalstairs}An illustration of a non-local order parameter for
measuring the phases of local symmetries directly. A segment is chosen from the chains and divided into three consecutive sections. The symmetry operations are applied to the middle sections. The left and right sections are permuted such that the endpoints with the same number were connected to each other before applying the permutation. The order parameter is then obtained by calculating the expression overlap with the four original replicas of the state. The labels $\Gamma$ and $\Lambda$ have been left out so that the figure does not become too busy; furthermore, there are
additional ``domain walls" at the other ends of the segments $L$ and $R$ which
are not shown.
}
\end{figure}
 Take $m$ identical copies of the given state, and place them side-by-side.  Take three successive long segments
$L$, $M$  and $R$. Now apply the  symmetries
to the middle segment and different permutations to the two ends (similar to the swap operator used in the case of time reversal symmetry). This causes the $U$
matrices at each of the ends of $M$ to get multiplied together
in two orders which allows one to detect the phase factor.
Fig. \ref{fig:musicalstairs} illustrates the order parameter for
the simplest case of $U_aU_bU_a^{-1}U_b^{-1}$ (where $\Sigma_a$ and $\Sigma_b$
commute).
The order parameter and its value is
\begin{widetext}
\begin{eqnarray}
\mathcal{S}&=&\langle \psi_m|\pi_{(1\ 2\dots m-1\ m)}(L)\Sigma^{a_1}(M_1)\Sigma^{a_2}(M_2)\dots \Sigma^{a_m}(M_m)\pi_{(k_1\ k_2\dots k_{m-1}\ k_m)}(R)|\psi_m\rangle\nonumber\\
&=& (\mathrm{tr}\Lambda^{2m})^4e^{i\phi}\label{eq:musicalsteps}.
\end{eqnarray}
\end{widetext}

Here $\pi_X(L)$ consists of permuting the left segments
of the chains according to the permutation $X$ (written as a cycle), $\Sigma^{a_k}(M_k)$ means to apply the symmetry
$a_k$ to the middle segment of the $k^\mathrm{th}$ chain, and
$\pi_X(R)$ indicates permuting the right segments. The wave function $|\psi_4\rangle$ is simple a product of $m$ replicas of the ground state.
To identify the phase factor in Eq. (\ref{eq:alphabet}), take
$m=8$ chains, and apply the symmetries $a,b,a^{-1},b^{-1},d,c,d^{-1},c^{-1}$ to
the legs of the ladder in the middle segment. Then apply the permutations  $(12345678)$ on the left segment and $(13245768)$ on the right one, in order to
get the $U$'s to cancel with their inverses on the right and to obtain $(U^{\vphantom{-1}}_aU^{\vphantom{-1}}_bU_a^{-1}U_b^{-1}U^{\vphantom{-1}}_dU^{\vphantom{-1}}_cU_d^{-1}U_c^{-1})^*$
on the left, which is the phase we want to find. (We could
also apply a bunch of two cycles on the right end to get the same
phase factor but multiplied by a different combination of Renyi entropies.)

This type of nonlocal order parameter distinguishes between all phases with a local
symmetry group. Time reversal and inversion symmetry phase
factors can be determined as in the previous section.  By
combining all these ideas together, it should also be possible to measure phase factors that arise from combining spatio-temporal symmetries with local symmetries.
(We have not yet worked out order parameters for groups that contain either inversion or time-reversal symmetry together with local symmetries, but it seems
likely to be possible.)
 \section{Conclusions}
A topological phase is a phase of matter which cannot be characterized by a local order parameter. Gapped phases in 1D systems can be completely characterized using tools related to projective representations of the symmetry groups. If the ground state is given in the form of a matrix-product state representation, the different topological phases can be directly detected by diagonalizing a  generalized transfer matrix (obtained from an overlap with the transformed matrix-product state). Based on this fact, we also introduced non-local order parameters which can be simply calculated using alternative representations. Such an order parameter could be determined
for a wave function using Monte Carlo simulations or possibly experimentally. 
The ordinary string order for the Haldane phase can be explained using
a selection rule that changes at the critical point: there are two types
of string orders (depending on the operators at the ends of the string),
one which vanishes in the trivial phase, and one which vanishes in the non-trivial one. This order parameter can be generalized to many cases, but not to all
groups.  An alternative order parameter that directly measures the ``projective phases" is required to distinguish among phases in general. Such an order parameter
work for all cases, including time reversal, inversion symmetry, and complicated
local symmetry groups.  Intriguingly, this parameter involves
measuring expectation values of string operators on multiple copies of
the system, even though these copies are uncorrelated.
\label{con}

\section*{Acknowledgment}
We thank Erez Berg and Masaki Oshikawa for useful discussions and the collaboration on a related project. A.M.T.  acknowledges the hospitality of the guest program of MPI-PKS Dresden.
\appendix
\label{app1}
\section{Example of a Phase that cannot be detected by String Order Selection Rules}
Let us assume that a state is invariant under a symmetry group $G$ which fulfills the group algebra $aba^{-1}b^{-1}=x=cdc^{-1}d^{-1}$.  (We simplify the expressions
by writing $a,b,\dots$ instead of $\Sigma_a,\Sigma_b,\dots$.) Then $U_a^{\vphantom{-1}}U_b^{\vphantom{-1}}U_a^{-1}U_b^{-1}U_d^{\vphantom{-1}}U_c^{\vphantom{-1}}U_d^{-1}U_c^{-1}=e^{i\phi}\mathds{1}$
is gauge-invariant and does not look easy to detect by using the
selection rule for string order from Sec. \ref{sec:so}.  
But this
is not completely obvious.  In fact, if $a$ and $b$ both commute with both $c$ and $d$, then
we can rearrange $aba^{-1}b^{-1}dcd^{-1}c^{-1}$
into $(ad)(bc)(ad)^{-1}(bc)^{-1}=1$. Thus $ad$ and $bc$
commute, allowing us to define a phase $\phi_{ad;bc}$ (in general, we define $\phi_{g_1,g_2}$ for commuting
symmetries $g_1,g_2$ as the
phase of $U_{g_1}U_{g_2}U_{g_1}^{-1}U_{g_2}^{-1}$). The phase $\phi$ can be expressed in terms of it.  In fact, rearranging
the expression for $e^{i\phi}$ and remembering to keep track of the phases that might arise from exchanging e.g. $U_a$
and $U_c$ (on account of the topological order), we find
that $\phi=\phi_{b;d}-\phi_{a;c}+\phi_{ad;bc}$, so $\phi$ reduces to simpler phase-factors which can all be determined
by using the string order selection rule.

More pairs of symmetries must be non-commuting to ensure that there is no way to simplify $\phi$. We will introduce
the following group algebra, with $c$ and $d$ relabeled $c_1$
and $c_2$,
\begin{eqnarray}
&&aba^{-1}b^{-1}=x\nonumber\\
&&c_1c_2c_1^{-1}c_2^{-1}=x\nonumber\\
&&ac_1a^{-1}c_1^{-1}=y_1;\ \ ac_2a^{-1}c_2^{-1}=y_2\nonumber\\
&&bc_1b^{-1}c_1^{-1}=bc_2b^{-1}c_2^{-1}=1.
\label{eq:group}
\end{eqnarray}
Besides the original symmetries $a,b,c_1,c_2,x$
we have introduced additional symmetries $y_1,y_2$ as the
commutators of some of them.
Aside from these conditions, we assume that all the generators
square to one. This group has $128$ elements.  The algebraic
relations defining it are complicated generalizations of the quaternion group; for example, the first equation corresponds to
the commutator of $i\sigma_x$ and $i\sigma_y$ being $-\mathds{1}$, which commutes with everything and squares to $\mathds{1}$ like $x$ does.

\emph{Example of ``undetectable" projective phase factors for this group:  }
We can write the elements in the group as products of $a,b,c_1,c_2$ and $x,y_1,y_2$.  The numbers
of $a,b,c_1,c_2$ factors are each the same modulo two no matter how we rearrange the factors. So
define $n_a(g)$ and $n_b(g)$ to be the number of $a$'s and $b$'s appearing modulo 2.
Assume the following projective phase factors:
\begin{equation}
U_{g_1}U_{g_2}=e^{i\rho(g_1,g_2)}U_{g_1g_2}=e^{i\pi n_a(g_2)n_b(g_1)}U_{g_1g_2}.\label{eq:leftright}
\end{equation}
(To create an example of an undetectable phase factor, we want a minus sign
to appear in the first of Eqs. (\ref{eq:group}) on replacing
the symmetries by their $U$ matrices, and we want this to be the only phase
factor that appears. These conditions lead to
Eq. (\ref{eq:leftright}).)
It is easy to check that this definition is consistent,
i.e., that $\rho(g_1g_2,g_3)+\rho(g_1,g_2)=\rho(g_2,g_3)+\rho(g_1,g_2g_3)$
The
calculation starts from the linearity of
$n_a$ and $n_b$, e.g.
$n_a(g_1g_2)=n_a(g_1)+n_a(g_2)\ (\mathrm{mod}\ 2)$.

Now let us show first that this is a non-trivial phase (by showing
that there is a non-trivial phase factor defined using four symmetries) and second that
this phase cannot be detected using a string order selection rule, because all the commuting pairs of elements
$g_1$ and $g_2$ also commute in the projective representation.  Hence this
is a nontrivial phase without any signature in the ordinary
string order.

\emph{The phase is  non-trivial} as $U^{\vphantom{-1}}_aU^{\vphantom{-1}}_bU_a^{-1}U_b^{-1}=-U^{\vphantom{-1}}_x$ while $U^{\vphantom{-1}}_{c_1}U^{\vphantom{-1}}_{c_2}U_{c_1}^{-1}U_{c_2}^{-1}=U^{\vphantom{-1}}_x$. Hence
the gauge-invariant phase factor from these four symmetries is $-1$. 

However, \emph{all \emph{two-symmetry} phase factors are trivial}.
To show this, we have to enumerate (at least partly) all the pairs $g,g'$ of symmetries that commute, and then
check that the $U$'s for them also commute.  Write $g=za^{n_a}b^{n_b}c_1^{n_1}c_2^{n_2}$ and $g'=z'a^{n'_a}b^{n'_b}c_1^{n'_1}c_2^{n'_2}$ where
the $n$'s are each $0$ or $1$ and the $z$'s are products of some combination of $x$, $y_1$ and $y_2$ (i.e.,
elements of the center of the group).  Since the commutators of any two of $a,b,c_1,c_2$ are in the center, 
 the commutator of $g$ and $g'$ can be calculated by evaluating the commutators of their factors
one pair at a time:
\begin{equation}
gg'g^{-1}g^{'-1}=x^{n_an_b'+n_bn_a'+n_1n_2'+n_2n_1'}y_1^{n_an_1'+n_1n_a'}y_2^{n_an_2'+n_2n_a'}.
\label{eq:alphagam}
\end{equation}
We want $g$ and $g'$ to commute, so the exponents of $x$,$y_1$, $y_2$
must be zero modulo 2. We will then want to calculate $U^{\vphantom{-1}}_gU^{\vphantom{-1}}_{g'}U_g^{-1}U_{g'}^{-1}$. which according to Eq. (\ref{eq:leftright})  is $(-1)^{n_an_b'+n_bn_a'}$.

In order for $g$ and $g'$ to commute, the exponents of the $y$'s must vanish:
\begin{equation}
n_an_1'+n_a'n_1\equiv n_an_2'+n_a'n_2\equiv 0\ \mathrm{mod}\ 2.
\label{eq:determinant}
\end{equation}
Consider all four possible combinations of values for $n_a$ and $n_a'$.  First, if $n_a=n_a'=0$,
then $U_g$ and $U_{g'}$ commute because $n_an_b'+n_bn_a'=0$.
Second, if $n_a=1$ and $n_a'=0$, then we must have $n_1'=n_2'=0$ by Eq. (\ref{eq:determinant}).  This implies
that $g'=z'$ or $z'b$. But this commutes with $g$ only in the former case (since $g$ has a factor of $a$ in it and
this does not commute with $b$), while there is a non-trivial phase factor only in the latter case.
The remaining two cases are similar.

Hence this group is an example where the regular string-order selection rule we described in Sec.~\ref{sec:so} does
 not help to identify this phase, while the alternative type of order in
Sec. \ref{sec:phasestring} does.

\section{Schur's Theorem on Projective Representations \label{app:schurmultiplier}}
Schur classified the types of projective representations
(which are also known as the ``Schur Multiplier"); the result\cite{Karpilovsky-1987}
shows that all one-dimensional phases, at least with local
symmetry groups, can be recognized using the order parameter of Sec. \ref{sec:phasestring}. 
The gauge-invariant phase-factors we have found involve products
such as $U^{\vphantom{-1}}_aU^{\vphantom{-1}}_bU_a^{-1}U_b^{-1}U^{\vphantom{-1}}_dU^{\vphantom{-1}}_cU_d^{-1}U_c^{-1}$ where $a$,$b$,$c$,$d$ are
generators of the group. This phase factor can be defined by listing
the sequence of group elements that have to be multiplied together:
$<a,b,a^{-1},b^{-1},d,c,d^{-1},c^{-1}>$, without actually multiplying them. It is
convenient to regard such sequences as forming a group (a ``free group"): to multiply
two sequences, juxtapose them and cancel elements with their inverses
when they meet each other. This structure is useful because it makes
it possible to break phase factors down to simpler ones (for
example, repeating the string just given twice does not give
a new phase factor, just the square of the original one).

The sequences that give a gauge-invariant phase
factor form a group, which Schur's theorem describes.
In general, let the symmetry group be $G$ and let
$x_1,\dots,x_k$ be a set of symmetries that generate it; call the
set of sequences of these generators the free group $F$. Let $[F,F]$
be the group generated by commutators of two elements of $F$. (Similarly,
one can define the commutator of any two subgroups $[A,B]$).
Consider also the set $R$ of sequences whose product is equal to the identity.
A sequence that lies in both these subgroups determines a gauge-invariant phase factor; that is, there is a function $e^{i\phi(\gamma)}$ defined
on $\gamma\in [F,F]\cap R$.
Because $\gamma$ is an element of $R$, it gives a phase when the corresponding 
$U$'s are multiplied
together, and these phases are invariant because elements of
$[F,F]$ are products of commutators, that is, they
have the form $<a_1,b_1,a_1^{-1},b_1^{-1},\dots ,a_l,b_l,a_l^{-1},b_l^{-1}>$ where
the $a$'s and $b$'s are various elements of $F$.
 The theorem of Schur states that all gauge-invariant phase factors are
contained in this function. In addition, the theorem finds all the conditions that have
to be satisfied by these phases (such as when one of the phases has
to be $\pm 1$); the general rule is that $\phi(\gamma)=0$
when $\gamma\in [R,F]$.

Hence, the classes of projective representations of a group are in one-to-one
correspondence with characters $e^{i\phi(\gamma)}$ on the group
$([F,F]\cap R)/[R,F]$. When $G$ is finite, this is a finite group, too.

For example, consider $Z_2\times Z_2$.
Let $a$ and $b$ be the two generators.
Then $x=<a,b,a^{-1},b^{-1}>$ is an element of $R$ because $a$ and $b$
commute as elements of the group $Z_2\times Z_2$. It is also an element of $[F,F]$,
so it defines a phase factor $e^{i\phi_{ab}}$.  Now the second
part of the theorem implies that this phase factor is equal to $\pm 1$.
To see this, we will show that $x^2\in [R,F]$, which implies
$(e^{i\phi(x)})^2=e^{i\phi(x^2)}=1$. That $x^2\in [R,F]$
is implied by the following relationship (where the commas represent multiplication
in the free group):
$x^2=<(x,a,x^{-1},a^{-1}),(a^2,b,a^{-2},b^{-1})>$. This is in $[R,F]$ because
$x$ and $a^2$ are both in $R$.

\end{document}